\documentclass[aps,prb,twocolumn]{revtex4}
\usepackage{graphicx,amsmath,amssymb}
\usepackage[usenames,dvipsnames]{pstricks}
\usepackage{epsfig}

\begin{document}

\title{Wave Packets and Scattering Probabilities}
\author{Travis Norsen}
\affiliation{Marlboro College \\ Marlboro, VT  05344}

\date{October 8, 2009}

\begin{abstract}
We propose a simple, intuitive alternative method of deriving the rule
for connecting asymptotic wave function amplitudes to scattering
probabilities.  This is illustrated using the standard example of a
1-D particle reflecting or transmitting from a potential step.  
\end{abstract}

\maketitle


The quantum mechanical calculation of scattering probabilities generally
requires two inputs:  (first) asymptotic wave function
amplitudes for outgoing
waves in various directions, and (second) a rule for how these 
amplitudes relate to the probabilities for a particle to scatter in
those various directions.  

Let us illustrate with the simplest possible example -- a
non-relativistic particle of mass $m$ in 1-D incident (from the left) 
on a potential step:
\begin{equation}
V(x) = V_0 \; \theta(x) = \left\{
\begin{array}{lll}
0 & {\mathrm{for}} & x<0 \\
V_0 & {\mathrm{for}} & x>0 \\
\end{array}
 \right.
.
\label{potential}
\end{equation}
The asymptotic wave function amplitudes can
be found by solving the time-independent Schr\"odinger equation for an
energy eigenstate of energy $E$ (and which has no incoming wave 
component on the right):
\begin{equation}
\psi_E(x) = \left\{
\begin{array}{lll}
A e^{i k x} + B e^{-i k x} & \text{for} & x<0 \\
C e^{i \kappa x} & \text{for} & x>0 \\
\end{array}
\right.
\label{wf}
\end{equation}
where $k^2 = 2 m E/\hbar^2$ and $\kappa^2 = k^2 - 2 m V_0 /
\hbar^2$. A standard calculation reveals that this is a solution of
Schr\"odinger's equation at the origin only if the the wave function
amplitudes for the reflected ($B$) and transmitted ($C$) waves satisfy
\begin{equation}
\frac{B}{A} = \frac{k - \kappa}{k + \kappa}
\label{BoverA}
\end{equation}
and
\begin{equation}
\frac{C}{A} = \frac{2 k}{k + \kappa}
\label{CoverA}
\end{equation}
respectively.

The usual method of converting these expressions into scattering
probabilities involves associating a probability current $j$ with each
of the three terms in Equation \eqref{wf}, and then arguing that the
reflection and transmission probabilities should be
given respectively by $R = |j_B|/j_A$ and $T = j_C / j_A$ 
(in what is hopefully a transparent notation).  
This approach hides some subtle and dubious aspects, though,
and the introduction of the new idea of probability current (often,
just to facilitate the connection between wave function amplitudes and
scattering probabilities) is an extra cognitive load on students.

We therefore propose here an alternative approach which is based on a
simple time-dependent picture of a (very wide) \emph{wave packet}
incident on, and
then scattering from, the potential step at $x=0$.  

\begin{figure}[t]
\centering
\scalebox{0.9}{
\scalebox{1} 
{
\begin{pspicture}(0,-3.025)(8.72,3.01)
\usefont{T1}{ptm}{m}{n}
\rput(6.601406,2.67){$V(x)$}
\usefont{T1}{ptm}{m}{n}
\rput(8.081407,0.85){$x$}
\psbezier[linewidth=0.05](5.1,-1.6)(4.8,1.6)(4.98,-2.888618)(4.7,-2.9)(4.42,-2.9113822)(4.7,-0.2886179)(4.4,-0.22)(4.1,-0.15138209)(4.4,-3.0)(4.1,-3.0)(3.8,-3.0)(4.1,-0.2)(3.8,-0.2)(3.5,-0.2)(3.8,-3.0)(3.5,-3.0)(3.2,-3.0)(3.5,-0.2)(3.2,-0.2)(2.9,-0.2)(3.2,-3.0)(2.9,-3.0)(2.6,-3.0)(2.9,-0.2)(2.6,-0.2)(2.3,-0.2)(2.6,-3.0)(2.3,-3.0)(2.0,-3.0)(2.3,-0.2)(2.0,-0.2)(1.7,-0.2)(2.0,-3.0)(1.7,-3.0)(1.4,-3.0)(1.7,-0.24)(1.4,-0.24)(1.1,-0.24)(1.3821588,-2.8975303)(1.1,-2.9)(0.81784123,-2.9024699)(1.1,1.3)(0.78,-1.6)
\usefont{T1}{ptm}{m}{n}
\rput(6.911406,-0.09){\psframebox*[framesep=0, boxsep=false,fillcolor=White] {$Re[\psi(x)]$}}
\psline[linewidth=0.04cm,linestyle=dashed,dash=0.16cm 0.16cm,tbarsize=0.07055555cm 5.0]{|-|*}(0.7,0.16)(5.16,0.16)
\psframe[linewidth=0.04,linecolor=White,dimen=outer,fillstyle=solid](2.94,0.42)(2.54,-0.08)
\usefont{T1}{ptm}{m}{n}
\rput(3.0714064,0.17){\psframebox*[framesep=0, boxsep=false,fillcolor=White] {$w_I \;$}}
\psline[linewidth=0.04cm,arrowsize=0.1529cm 1.0,arrowlength=1.31,arrowinset=0.2]{->}(0.040699005,-1.6)(8.7,-1.6)
\psline[linewidth=0.05](0.0,1.0)(5.98,1.04)(5.98,2.24)(8.58,2.24)
\psline[linewidth=0.02,arrowsize=0.1529cm 2.0,arrowlength=1.4,arrowinset=0.2]{<->}(6.0,3.0)(5.98,1.04)(8.68,1.04)
\psline[linewidth=0.024cm,arrowsize=0.1529cm 2.0,arrowlength=1.4,arrowinset=0.2]{->}(6.0,-1.6)(6.0,0.4)
\usefont{T1}{ptm}{m}{n}
\rput(8.181406,-1.85){$x$}
\psline[linewidth=0.05cm](0.0,-1.6)(0.82,-1.6)
\psline[linewidth=0.05cm](5.1,-1.62)(8.5,-1.6)
\psline[linewidth=0.02cm](5.98,2.04)(6.0,0.58)
\psline[linewidth=0.02cm](6.0,-0.06)(6.0,-2.02)
\end{pspicture} 
}
}
\caption{A generic wave packet (with approximately-constant amplitude
  over most of its width, $w_I$) is incident on the step potential's
  scattering center at $x=0$.}
\label{fig1}
\end{figure}
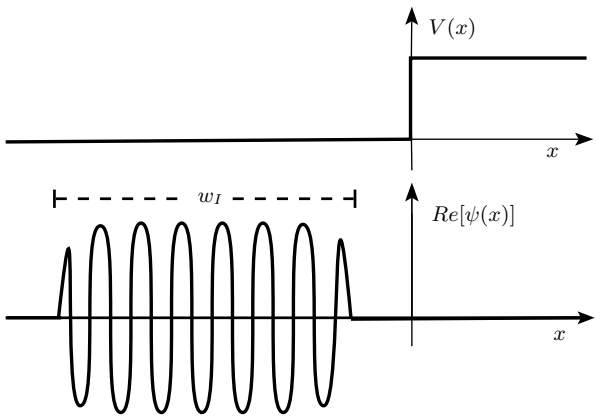

Consider a wave packet approaching the scattering center at $x=0$ for
the potential defined in Equation \eqref{potential}, as indicated in
Figure~\ref{fig1}.
Assume the packet has an almost-exactly constant amplitude ($A$)
and wavelength ($\lambda_0 = 2 \pi / k_0$)
in the region (of width $w_I$) where the amplitude is
non-vanishing, as
shown in the Figure.  Thus, where the amplitude is non-zero, the
packet will at each moment be well-approximated by a plane wave:
\begin{equation}
\psi = A \, e^{i k_0 x}.
\end{equation}
We may assume
this incident packet is normalized, so that $|A|^2 w_I = 1$.

What happens as the packet approaches and then interacts with the
potential step at $x=0$?  We
assume that the inevitable spreading of the wave packet is negligible
on the relevant timescales, so that the packet retains its overall
shape as it approaches the scattering center, moving at the group
velocity corresponding to the central wave number for the region $x<0$:
\begin{equation}
v_{g}^{<} = \frac{\hbar k_0}{m}.
\label{vg<}
\end{equation}
Now suppose the leading edge arrives at the scattering center 
at time $t_1$, so that the trailing edge arrives at $t_2$ satisfying
\begin{equation}
t_2 - t_1 = w_I / v_{g}^{<} = w_I m / \hbar k_0 .
\end{equation}
For intermediate
times, $t_1 < t < t_2$, we will have, in some (initially
small, then bigger, then small again) region surrounding $x=0$,
essentially the situation described in Equation \eqref{wf}.
In particular, the same 
relations mentioned earlier for the relative amplitudes
of these three pieces -- Equations \eqref{BoverA} and \eqref{CoverA}
-- will (under the conditions already described) still apply.

In this dynamical picture, however, we interpret Equations
\eqref{BoverA} and \eqref{CoverA} as giving the amplitudes of
reflected and transmitted wave packets, whose leading and trailing
edges are created respectively at $t_1$ and $t_2$.  We may then
calculate the reflection and transmission probabilities as follows.

Consider first the reflected packet.  The probability of reflection,
$R$, is just its total integrated probability density --
which here will be its intensity $|B|^2$ times its width $w_R$.  But
the width of the reflected packet will be the same as the
width of the incident packet:  because these two packets both
propagate in the same region, they have the same group velocity, so
the leading edge of the reflected packet will be
a distance $w_I$ to the left of $x=0$ when the trailing edge of the
reflected packet is formed. Thus, we have
\begin{equation}
R = w_R |B|^2 = w_I |B|^2 = \left| \frac{B}{A} \right|^2
\label{R}
\end{equation}
where we have used the normalization condition for the incident packet.

Similarly, the total probability associated with the transmitted wave
will be its intensity $|C|^2$ times its width $w_T$.  But $w_T$ will
be \emph{smaller} than $w_I$ because the group velocity on the right is
smaller than on the left.  In particular: the leading edge of the
transmitted packet is created at $t_1$; the trailing edge is created
at $t_2$; and between these two times the leading edge
will be moving to the right at speed
\begin{equation}
v_{g}^{>} = \frac{\hbar \kappa_0}{m}
\end{equation}
where $\kappa_0^2 = k_0^2 - 2 m V_0 / \hbar^2$ is the (central) wave
number associated with the transmitted packet.  Thus, the width of the
transmitted packet -- the distance between its leading and trailing
edges -- is
\begin{equation}
w_T = v_{g}^{>} \; (t_2 - t_1) = \frac{\kappa_0}{k_0} w_I
\end{equation}
and so the transmission probability is
\begin{equation}
T = w_T |C|^2 = \frac{\kappa_0}{k_0} \left| \frac{C}{A} \right|^2
\label{T}
\end{equation}
in agreement with the usual expression derived using probability
current ratios.

In addition to avoiding the need to discuss probability currents and
providing a more intuitive understanding of the
perhaps-puzzling factor of $\kappa_0/k_0$ in Equation \eqref{T}, the
approach outlined here has several advantages associated with the
explicit bringing-in of the time-dependent dynamical picture of wave
packet scattering.  The obvious point here is that this picture will
allow students to understand their \emph{calculations} of ``scattering
probabilities'' in terms of a recognizable physical process of
\emph{scattering}.  (In our approach, Equation \eqref{wf} is taken as
a description which applies only near the origin and only for a
certain period of time -- prior to which there was an incident packet,
and after which there are reflected and transmitted packets.  
Without this background context, it is
actually non-sensical to take Equation \eqref{wf} as a description of
a \emph{scattering} particle.)  The less obvious point is that having
the dynamical wave packet in mind encourages
one to notice and conceptualize the several overlapping
\emph{approximations} which are embodied in Equations \eqref{R} and
\eqref{T}.  (In particular, and given that the rigorous quantum
mechanical description of a scattering particle requires a wave
packet, the packet must be \emph{smooth} and \emph{very wide} compared
to other length scales in the problem, such as $\lambda_0$ and any
finite spatial structure in the scattering center.  The presence of
such assumptions in the standard derivation is completely obscure.)  

The only apparent disadvantage of this alternative approach is that, while
eliminating the need to introduce probability currents, it introduces
a need to confront group velocities.  But given the overall conceptual
advantages associated with thinking of scattering as a dynamical
process involving wave
packets, we think this price is more than outweighed by the
corresponding gains.  

It should also be noted that this alternative approach to  deriving
the relation between asymptotic wave function amplitudes and
scattering probabilities is completely general.  It would apply not
only to more complicated examples of 1-D scattering, but 3-D
scattering as well (though in 3-D it is less common for the group
velocities in different directions to differ).

\vspace{.5 cm}

\emph{Acknowledgements:}
Thanks to Sam McKagan, Josh Lande, and Mike Dubson for helpful
discussions and contributions to an earlier, longer version of the
paper.\footnote{T. Norsen, J. Lande, S.B. McKagan, 
``How and why to think about scattering in terms of wave packets 
instead of plane waves,''  arXiv:0808.3566v2  }

\end{document}